\documentclass[danish,a4paper,10pt,
twocolumn,
amsmath,amssymb,floatfix,superscriptaddress,
aps,
prl,showpacs
]{revtex4-1}
\pdfoutput=1
\usepackage{graphicx}
\usepackage[english]{babel}
\usepackage{bm}
\usepackage{braket}
\usepackage{float}
\usepackage{mathtools}
\usepackage{dcolumn}
\usepackage[draft]{fixme}
\usepackage{color}
\usepackage{lipsum}
\definecolor{red}{rgb}{1,0,0}
\definecolor{blue}{rgb}{0,0,1}
\usepackage{stmaryrd}
\usepackage{braket}
\usepackage{bbold}
\usepackage{amsmath}
\usepackage{amsfonts}
\usepackage{hyperref}
\usepackage{amssymb}
\usepackage{upgreek}
\usepackage{makecell}
\usepackage{wrapfig}
\usepackage{xr-hyper}
\usepackage{xcolor}
\usepackage{hyperref}

\definecolor{knote}{RGB}{195,75,227}

\begin{document}

\title{Revealing the strokes of autonomous quantum heat engines with work and heat fluctuations}
\author{Kat\'erina~Verteletsky}
\email{kverteletsky@phys.au.dk}
\affiliation{Department of Physics and Astronomy, Aarhus University, 8000 Aarhus C, Denmark}
\author{Klaus~M{\o}lmer}
\email{moelmer@phys.au.dk}
\affiliation{Department of Physics and Astronomy, Aarhus University, 8000 Aarhus C, Denmark}

\date \today
\begin{abstract}
We analyze an autonomous thermoelectric engine composed of two superconducting qubits coupled to separate heat baths and connected by a Josephson junction. Work and heat are process quantities and not observables of the engine quantum system, but their rates can be derived from the steady state expectation value of appropriate system observables, and their fluctuations are given by correlation functions determined by the master equation and quantum regression theorem of open quantum systems. Underneath the constant steady state of the system, the temporal correlation functions reveal a cyclical, dynamical transfer of energy---the strokes of the engine.
\end{abstract}

\maketitle

\begin{figure}[h!]
\includegraphics[width=0.48\textwidth]{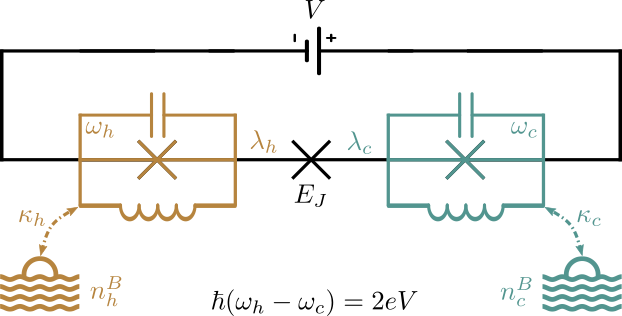}
\caption{\label{fig:2qubitengine_2heatbeaths}Schematic representation of a two-qubit engine powered by heat baths and performing work by resonant tunneling of Cooper pairs against a constant bias voltage $V$. Excitation transfer between the qubits is mediated by a tunneling current with Josephson oscillations tuned to the difference $2eV=\hbar(\omega_h- \omega_c)$ between the qubit excitation energies. The qubits are coupled to the Josephson junction with constants $\lambda_{h,c}$ and to bosonic heat baths with average excitation numbers $n_{h,c}^B$ with rates $\kappa_{h,c}$.}
\end{figure}

\paragraph{\label{main:intro}Introduction.}
Quantum thermodynamics is a research field that both addresses practical engineering issues arising with the miniaturization of physical machines \cite{heatcomputerchips1,heatcomputerchips2,qthreviewanders} and investigates the more fundamental relationship between statistical physics and (quantum) information science \cite{jarzynski,qtmworkklaus,workmeasurement,workfromcorr,thcostcorrelations,qtmjarzynski,landauer,qtmlandauer,qthreviewanders,qthreviewkosloff}.
These investigations have been accompanied by theoretical proposals for actual quantum machines, following the early works of Alicki~\cite{first-alicki} and Kosloff~\cite{first-kosloff}, such as quantum heat pumps~\cite{qtmcarnotcycle}, noise-driven quantum absorption refrigerators~\cite{qtmabsorptionfridge,qtmfridgeimplemexamples} and thermal valves~\cite{pekola}.
Experimental realizations of quantum thermal machines have been achieved with trapped ions~\cite{expfridge3ions,expheatengine1ion}, and a circuit QED thermoelectric engine has been proposed in~\cite{refJ}, where an electric current is driven by excitation transfer between superconducting resonators in contact with separate heat baths.\\
As in classical thermodynamics, work and heat are not state functions but process quantities, defined by the exchange of energy between the engine and its environment. For transient processes
of finite duration, this has led to definitions of heat and work that refer explicitly to the dissipative and Hamiltonian parts of the system evolution~\cite{qtm-meas-th-auffeves,qtmworkfluctuations}, while generalizations to the quantum regime of classical fluctuation theorems~\cite{bk1,jarzynski,jarzynski2,crooks1,crooks2,hummer-szabo,alberto} have employed measurement statistics~\cite{fluctth-measstat-hanggi,campisi-hanggi-talkner} and quantum trajectory dynamics~\cite{qtmfluctth-auffeves,flucth-qtmjump-nissila,qtmjump-pekola}.
In this article we study an autonomous heat engine operated out of thermal equilibrium---due to its constant coupling to baths with different temperatures---for which we can derive a steady state density matrix. For this system, work and heat are exchanged with the environment with mean rates that we can express by steady state expectation values of suitable system observables. Fluctuations in the integrated work and heat over finite time intervals are not simply the variances of the same observables but employ more complex quantum correlation functions. In particular, the transfer of heat into the cold bath is equivalent to the process of spontaneous emission from a quantum light source, and its temporal correlations thus follow from Glauber's photodetection theory~\cite{glauber1} in quantum optics.
\paragraph{\label{main:system}System Hamiltonian and master equation.}
Our analysis and calculations may be applied to many systems, but we shall for concreteness study the minimal quantum heat engine with two qubits in contact with separate heat baths originally proposed by Linden, Popescu, and Skrzypczyk~\cite{popescu}. In~\cite{popescu} the excitation of an ancillary ladder system accounted for the work done by the engine, while an experimental implementation with oscillators instead of qubits and a DC-voltage biased Josephson junction instead of the ladder system has been proposed in~\cite{refJ}. We refer to a similar architecture shown schematically in Fig.~\ref{fig:2qubitengine_2heatbeaths}, where we have supplemented the oscillators in~\cite{refJ} with non-linear elements, and we assume restriction of the dynamics to the two lowest oscillator levels to effectively implement Linden, Popescu, and Skrzypczyk's two-qubit model.

Following the same procedure as Hofer \textit{et al.} in Ref.~\cite{refJ}, we obtain the two qubit Hamiltonian~\cite{refJref40}:
\begin{eqnarray}
\hat{H} & = & \hbar \omega_h \hat{\sigma}^+_h \hat{\sigma}^-_h + \hbar \omega_c \hat{\sigma}^+_c \hat{\sigma}^-_c  \\
& \quad \quad & - E_J \cos[2eVt + 2\lambda_h (\hat{\sigma}^+_h + \hat{\sigma}^-_h) + 2\lambda_c (\hat{\sigma}^+_c + \hat{\sigma}^-_c)]\nonumber
\label{eqn:theory:A.1:josephsonham}
\end{eqnarray}
where $\hat{\sigma}^+_h$ ($\hat{\sigma}^+_c$) and $\hat{\sigma}^-_h$ ($\hat{\sigma}^-_c$) are the Pauli spin operators exciting and deexciting the hot (cold) qubit with oscillation frequency $\omega_h$ (respectively $\omega_c$),  $E_J$ the energy of the junction, $V$ the bias voltage across the junction, and $\lambda_h$ ($\lambda_c$) is the coupling constant of the hot (cold) qubit to the junction. Note that the time-dependent term arises from the Josephson interaction with a constant voltage bias. Passing to a rotating frame at the Josephson oscillation frequency and using the Baker-Campbell-Hausdorff formula, we employ the rotating wave approximation and retain only terms representing resonant excitation transfer between the two qubits:
 \begin{eqnarray}
\hat{H}_{RWA} = \frac{E_J}{2} \sin(2\lambda_h)\sin(2 \lambda_c) (\hat{\sigma}^-_h \hat{\sigma}^+_c + \hat{\sigma}^+_h \hat{\sigma}^-_c).
\label{eqn:theory:A.1:rwaham}
\end{eqnarray}
%
%
The resonance condition, $\hbar(\omega_h-\omega_c)=2eV$, ensures that the transfer of a quantum of excitation is accompanied by the passage of a Cooper pair of charge $2e$ across the junction, i.e., against the voltage difference $V$~\cite{potts-cooper}. A sustained excitation transfer is possible because the qubits are coupled to their respective heat baths, causing excitation of the hot qubit and dissipation of the energy transferred to the cold qubit. While these elements of the dynamics are analogous to the heat addition and removal steps in classical heat cycles, the qubit cycle is not enforced by any external agent. The system is described by a density matrix  with a time independent master equation and it reaches a constant steady state.

The density operator for the two-qubit system can be expressed in the joint product basis $\ket{i_h}\otimes \ket{k_c}$ of the hot and cold qubits.
We assume the validity of the Born-Markov approximation and obtain the conventional Gorini-Kossakowski-Lindblad-Sudarshan (GKLS) master equation with coupling to the thermal baths:
\begin{eqnarray}
\frac{d \hat{\rho}(t)}{dt} = -i [\hat{H}_{RWA},\hat{\rho}(t) ] + \mathcal{L}_B\hat{\rho}(t),
\label{eqn:theory:A.1:me}
\end{eqnarray}
where the dissipative interaction with the heat baths is represented by:
\small
\begin{eqnarray}
 \mathcal{L}_B \hat{\rho}(t) &=& \sum_{\alpha=c,h}  \kappa_\alpha (n_\alpha^B +1) \nonumber \\
&& \quad \quad \quad \times \left [ \hat{\sigma}^-_\alpha \hat{\rho}(t) \hat{\sigma}^+_\alpha - \frac{1}{2} (\hat{\sigma}^+_\alpha \hat{\sigma}^-_\alpha \hat{\rho}(t) + \hat{\rho}(t) \hat{ \sigma}^+_\alpha \hat{\sigma}^-_\alpha) \right ] \nonumber \\
&& \quad \quad  + \kappa_\alpha (n_\alpha^B) \nonumber \\
&& \quad \quad \quad \times \left [ \hat{\sigma}^+_\alpha \hat{\rho}(t) \hat{\sigma}^-_\alpha - \frac{1}{2}( \hat{\sigma}^-_\alpha \hat{\sigma}^+_\alpha \hat{\rho}(t) + \hat{\rho}(t) \hat{\sigma}^-_\alpha \hat{\sigma}^+_\alpha) \right ]. \nonumber \\
\label{eqn:theory:dissipativeterms}
\end{eqnarray}
\normalsize
This, so-called, local master equation treats the interaction of the physical components with the respective heat baths separately and is valid for coupling strengths much weaker than the qubit excitation frequencies ($E_J \sin(2\lambda_h)\sin(2 \lambda_c) \ll \omega_h,\omega_c$). For an intermediate range of coupling strengths, diagonalizing the Hamiltonian first and coupling the baths to global observables yields similar results~\cite{localglobal-potts,localglobal-huelga,localglobal-correa}, while for coupling strengths beyond the regime explored in this work, the system would interact with bath degrees of freedom at the strongly modified dressed state energies, and the local master equation would not appropriately describe its dynamics.
The steady state solution of Eq.~\eqref{eqn:theory:A.1:me} is found by setting $\frac{d \hat{\rho}(t)}{dt} = 0$.
\paragraph{\label{main:workheat}Average work and heat.}
The work performed by the engine, e.g., to charge a battery providing the voltage $V$, is readily defined as the product of the voltage and the electrical charge transferred through the Josephson junction. Since a charge of $2e$ is transferred with each excitation transfer to the cold qubit, the current through the junction is given by $2e$
multiplied with the rate of change of the cold qubit excited state population
 due to the commutator with the Hamiltonian (\ref{eqn:theory:A.1:rwaham}). This motivates our introduction of the current operator:
\begin{eqnarray}
\hat{I} &=& \frac{2e}{i\hbar} [\hat{\sigma}^+_c \hat{\sigma}^-_c,\hat{H}_{RWA}]\\ \nonumber
 &=&-  i eE_J \sin(2\lambda_h)\sin(2 \lambda_c) (\hat{\sigma}^-_h \hat{\sigma}^+_c - \hat{\sigma}^+_h \hat{\sigma}^-_c).
\label{eqn:theory:B:IRWA}
\end{eqnarray}

\begin{figure}[h!]
\begin{center}
\includegraphics[width=0.48\textwidth]{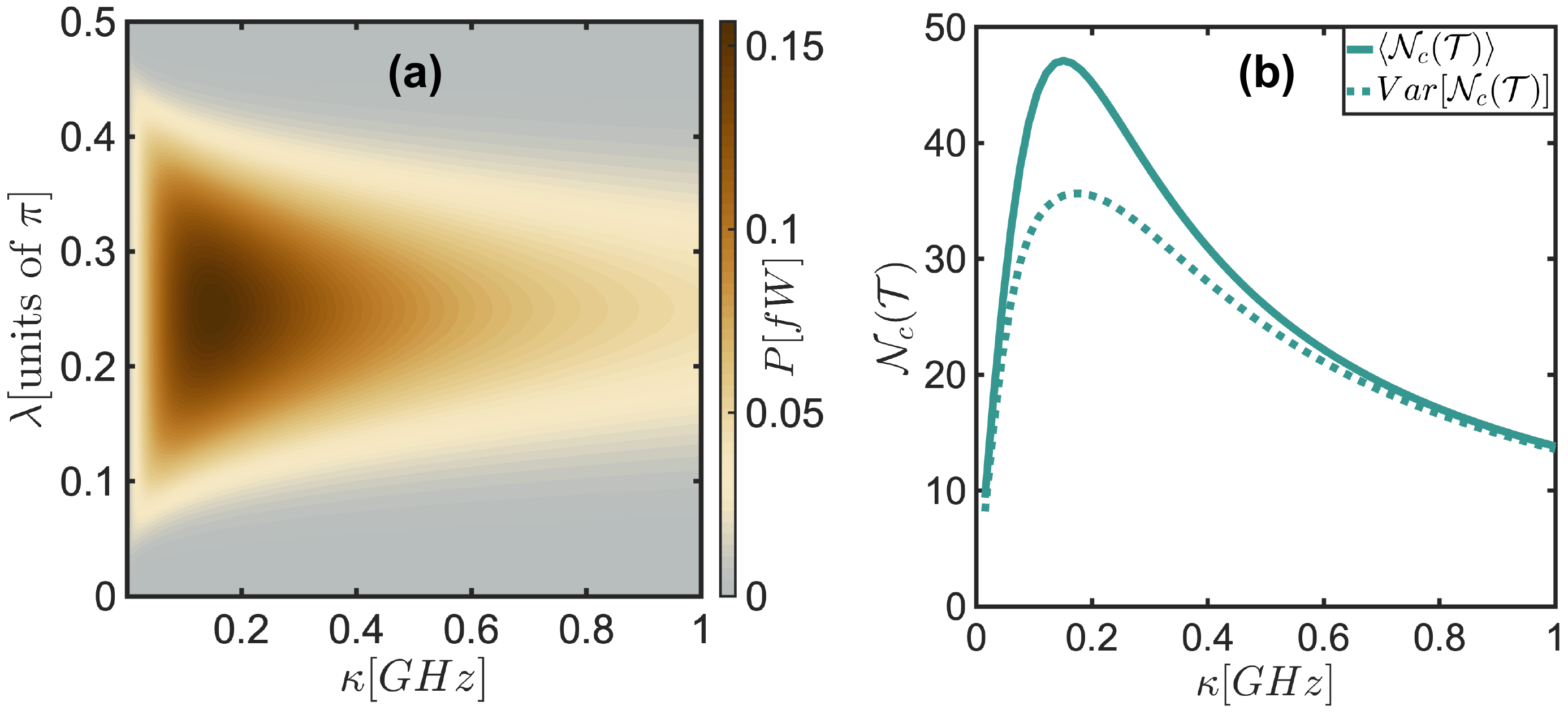}
\caption{\label{fig:Pklvarmean}Panel (a) shows the steady state output power of the engine as a function of the bath coupling constant $\kappa$ and qubit coupling to the Josephson junction $\lambda$ ($=\lambda_h=\lambda_c$). The system parameters are %
based on Ref.~\cite{refJ}: $E_J=2\pi \times 0.3$GHz, $\omega_h=2\pi \times 13.5$GHz, $\omega_c=2\pi \times 3.0$GHz, $n_h^B=1.5$, $n_c^B=0$. The maximum power, $P_{max}^{\kappa,\lambda}=0.16$fW, is reached for $\lambda=\pi/4$ and $\kappa= 2 \pi \times 0.151$GHz. Panel (b) shows the mean value and fluctuations of the number of quanta $\mathcal{N}_c(\mathcal{T})=Q_c(\mathcal{T})/(\hbar \omega_c)$, transferred to the cold bath during a time interval, $\mathcal{T}=100 \times (1/E_J)$, as function of $\kappa$ with the same system parameters as in panel (a) and $\lambda=\pi/4$. The excitation transfer count exhibits a sub-Poissonian behavior.
}
\end{center}
\end{figure}

Restricting ourselves to the case $\kappa_h = \kappa_c = \kappa$, we derive an analytical expression for the steady state density matrix and hence for the steady state expectation value of the current:
\begin{eqnarray}
\langle \hat{I} \rangle_{ss} &=&\frac{e}{(n_h^B+n_c^B+1)} \nonumber \\
&& \times \frac{n_h^B - n_c^B}{\left [ \frac{1}{\kappa}+\frac{\kappa}{(E_J')^2} (2 n_c^B + 1)(2 n_h^B + 1) \right ]  }  \nonumber\\
\label{eqn:theory:B:IRWAss}
\end{eqnarray}
where $E_J'=E_J \sin(2\lambda_h)\sin(2 \lambda_c)$. Using this expression, we can readily evaluate the average output power $P =\langle \hat{I} \rangle_{ss} V$.
The presence of the term $(n_h^B-n_c^B)$ shows that the directionality of the engine is governed by the bath occupation number difference (with a negative current obtained in the case $n_h^B<n_c^B$). As a function of the coupling strength $\kappa$, the output power reaches a maximum value and decreases for stronger couplings to the baths which suppress the qubit coherence responsible for the excitation transfer (see Fig.~\ref{fig:Pklvarmean}(a). When $\kappa < E_J \ll \omega_h, \omega_c$, the global master equation would lead to results similar to the ones shown in the figure~\cite{localglobal-potts}.

The heat transfer between the qubits and their respective baths is determined by the rate of change of the excited state populations due to the dissipative terms in the master equation (\ref{eqn:theory:A.1:me},\ref{eqn:theory:dissipativeterms}). If we restrict our analysis to the case where the cold bath has vanishing temperature $n^B_c =0$, the mean rate of quanta dissipated by the cold qubit equals $\kappa\langle\hat{\sigma}^+_c \hat{\sigma}^-_c\rangle$  and the corresponding power dissipated as heat reads $\hbar\omega_c \kappa \langle \hat{\sigma}^+_c \hat{\sigma}^-_c\rangle$.

In steady state, the integrated work and heat transferred to the cold bath over a time interval of duration $\mathcal{T}$ take the mean values
\begin{eqnarray}
W(\mathcal{T}) = \langle \hat{I} \rangle_{ss}V \mathcal{T},
\label{eqn:theory:C:meanworkintfy}
\end{eqnarray}
and
\begin{eqnarray}
Q_c(\mathcal{T}) =  \hbar\omega_c \langle \hat{\sigma}^+_c \hat{\sigma}^-_c\rangle \kappa \mathcal{T},
\label{eqn:theory:C:meanheatintfy}
\end{eqnarray}
respectively.

Since the populations of the qubit states are constant in steady state, their rate of change due to dissipation is the exact opposite of their rate of change due to the Hamiltonian term in the master equation. The rate of loss of quanta from the cold qubit, hence, equals both the rate of transfer of Cooper pairs across the junction and the net rate of excitation of the hot qubit (subtracting the excitation and deexcitation rates by the coupling to the hot bath). The total energy is conserved and the total transfer from the hot bath equals the sum of the work and the heat delivered to the cold bath.
\begin{figure*}[ht]
\includegraphics[width=1\textwidth]{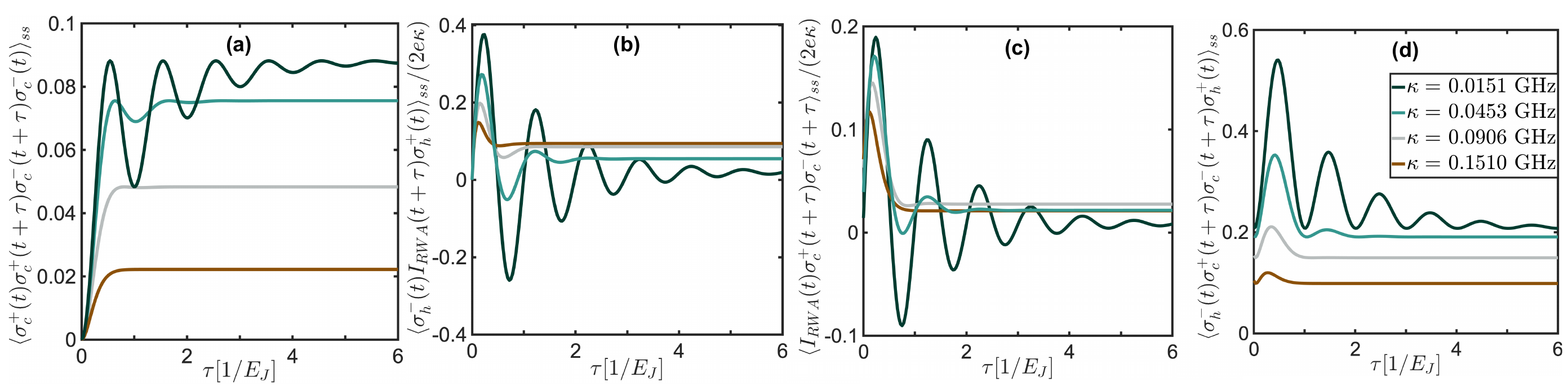}
\caption{\label{fig:corrbathsourkl2}Two-time correlations calculated for different values of the coupling $\kappa$ to the baths---see common legend in panel (d). The system is taken in steady state at the early time $t$, and $\tau$ is given in units of the junction oscillation frequency $1/E_J$. The other parameters are the same as in Fig.~\ref{fig:Pklvarmean}b. (a) Correlations for the heat emission into the cold bath. (b) Correlations between the heat extraction from the hot bath and the current through the junction. (c) Correlations between the current through the junction and the heat emission into the cold bath. (d) Correlations between the heat extraction from the hot bath and the heat emission into the cold bath.}
\end{figure*}

\paragraph{\label{main:fluctuations}Work and heat fluctuations.}
Work and heat are defined by the accumulated effects of coherent and incoherent processes involving the engine quantum system and its environment. In cyclic engines, with sequential interactions with heat baths and time-dependent Hamiltonians, procedures to characterize the work and heat and their fluctuations at each step have been subject of recent study~\cite{alberto,fluctth-measstat-hanggi,campisi-hanggi-talkner,qtmfluctth-auffeves}. Our situation is different as we deal with a continuous and simultaneous exchange of heat and work, and while the rates of these processes can be expressed in terms of mean values of system observables $\hat{I}$ and $\hat{\sigma}^+_c \hat{\sigma}^-_c$, their fluctuations are not given by the variances of the same operators in steady state.

To evaluate the work done by the engine, we may formally introduce the time integral of the current operator over a time interval of duration $\mathcal{T}$:
\begin{eqnarray}
\hat{W}(\mathcal{T}) = V \int_0^{\mathcal{T}}  \hat{I}(t)  dt.
\label{eqn:theory:C:acccharge}
\end{eqnarray}
This expression is written as the integral of the time dependent current, which is an operator in the Heisenberg picture. If we assume the system is in steady state throughout the time interval, the evaluation of its mean value can be done in the Schr{\"o}dinger picture with the time dependence transferred to the (constant) steady state density matrix, and we recover Eq.(\ref{eqn:theory:C:meanworkintfy}).

The operator expression \eqref{eqn:theory:C:acccharge} allows us to use the customary definition of variances for quantum observables, $\text{Var}[W(\mathcal{T})] = \langle \hat{W}(\mathcal{T})^2 \rangle - \langle \hat{W}(\mathcal{T})\rangle^2$. The first term in this equation can be written explicitly as:
\begin{eqnarray}
\langle \hat{W}(\mathcal{T})^2 \rangle
& = &  V^2 \left \langle  \left ( \int_0^{\mathcal{T}} \hat{I}(t) dt \right )  \left ( \int_0^{\mathcal{T}} \hat{I}(t') dt' \right ) \right \rangle  \nonumber \\
& = &   V^2 \int_0^{\mathcal{T}}\int_0^{\mathcal{T}} \langle \hat{I}(t)  \hat{I}(t')  \rangle  dt dt'.
\label{eqn:theory:C:q2}
\end{eqnarray}
According to the master equation (\ref{eqn:theory:A.1:me}) and the quantum regression theorem~\cite{quantumnoise-gardinerzoller}, the correlation function $\langle \hat{I}(t)  \hat{I}(t')  \rangle$ factorizes as $\langle   \hat{I} \rangle_{ss}^2$ for  $t,t',|t-t'|$ larger than a few $\kappa^{-1}$. Assuming ${\mathcal{T}} \gg \kappa^{-1}$, we may thus formally rewrite the double integral as  ${\mathcal{T}}$ multiplied with the integral of the steady state correlation function over the time difference $\uptau=t'-t$, for which the upper limit can be taken to infinity. This yields
\small
\begin{eqnarray}
\text{Var}[W(\mathcal{T})]_{\infty}  =  2 \mathcal{T} V^2 \int_0^{\infty}  \left [  \mathcal{R}e \left ( \langle \hat{I}(t)  \hat{I}(t+\uptau)\rangle_{ss} \right ) - \langle   \hat{I} \rangle_{ss}^2  \right ] d\uptau,  \nonumber\\
\label{eqn:theory:C:varwork2}
\end{eqnarray}
\normalsize
which is the constant-voltage equivalent of Eq. (25) in~\cite{new-prb}. The two-time correlation function is readily evaluated with the quantum regression theorem~\cite{quantumnoise-gardinerzoller}.\\
\indent The derivation of the heat fluctuations is conducted in a different manner, as the emission of quanta into the cold bath is an incoherent process governed by a rate. The number of quanta emitted into the cold bath during the time $\mathcal{T}$ is therefore a stochastic variable. The characteristic function of the number distribution of these quanta has been derived using master equations with counting fields in, e.g.,~\cite{new-pre,new-njp}, but in this article we adopt a more direct approach following the quantum optical description of photon counting processes.\\
\indent We may simulate the continuous detection of the arrival of quanta in the cold bath and assess the counting statistics of such a hypothetical experiment by a quantum jump stochastic master equation~\cite{mcwfprl,carmichaelm18}. The physical process of photodetection crucially relies on the excitation of an electron in the detector by the absorption of a photon from the incident quantized field. The probability of two such events at time $t$ and $t'$ thus relies on the conditional evolution of the field after the first absorption event. \\
\indent In the Heisenberg picture, the field annihilation operator can be expressed by the emitter lowering operator, and following the seminal work by Glauber~\cite{glauber1}, the two-click detection event occurs with a probability proportional to the normal- and time-ordered expression (for $ t'\ge t$): $G^2(t,t') = \kappa^2 \langle \hat{\sigma}_c^+(t)\hat{\sigma}_c^+(t')  \hat{\sigma}_c^-(t')\hat{\sigma}_c^-(t)  \rangle$. \\
\indent Assuming steady state at the early time $t$, $G^2(t,t')$ can be evaluated by use of the quantum regression theorem~\cite{quantumnoise-gardinerzoller}. Similar to the contribution of the current correlations to the work variance, the two-time correlation function for the detection of quanta arriving into the cold bath  provides the steady state variance of the total number of detection events in any given time interval~\cite{mandel-wolf-ch14,aleksander}. Unlike the calculation of work fluctuations, which were not associated with a measurement process, heat fluctuations are intimately related to the random counting statistics, and the way that detection disturbs the steady state and induces conditional transient dynamics. Indeed, the evaluation of $G^2(t,t')$ consists in applying, at time $t$, the cold qubit lowering operator on the steady state density matrix and subsequently propagating the resulting state by the master equation to evaluate the conditional emission rate at the later time $t'$.
The corresponding variance of the heat dissipated to the cold bath is proportional to the variance of the number of quanta emitted into the cold bath acting as a photodetector, and for long time intervals $\mathcal{T}$ it is given by~\cite{mandel-wolf-ch14,aleksander}:
\onecolumngrid
\small
\begin{eqnarray}
\text{Var}[Q_c(\mathcal{T})]_{\infty}
= (\hbar\omega_c)^2 \left (  2  \kappa^2 \mathcal{T}   \int_0^{\infty} \left [ \langle \hat{\sigma}_c^+(t)\hat{\sigma}_c^+(t+\uptau)  \hat{\sigma}_c^-(t+\uptau)\hat{\sigma}_c^-(t)  \rangle_{ss}  -  \langle \hat{\sigma}_c^+(t)  \hat{\sigma}_c^-(t) \rangle_{ss}^2 \right ] d\uptau
+   \kappa \mathcal{T} \langle \hat{\sigma}_c^+(t)  \hat{\sigma}_c^-(t) \rangle_{ss} \right ).
\label{eqn:theory:C:varnc2}
\end{eqnarray}
\normalsize
\twocolumngrid

The integrated number of quanta emitted into the cold bath, $\mathcal{N}_c(\mathcal{T})=Q_c(\mathcal{T})/(\hbar \omega_c)$, exhibits a sub-Poissonian behavior characteristic of photon counting from a two-level emitter~\cite{antibunching-kimble-mandel} (see Fig.~\ref{fig:Pklvarmean}(b). Despite their different origin in coherent and dissipative processes, and their quantitative properties being associated with direct operator integrals and measurement processes, respectively, the work produced by the machine and the heat transferred to the cold bath have equivalent mean values up to the ratio of the microscopic excitation energies, $\hbar \omega_c/2eV$. Their variances are also equivalent up to the corresponding factor $(\hbar \omega_c/2eV)^2$.
%
\paragraph{\label{main:strokes}Strokes in the engine dynamics.}
In addition to our characterization of the steady state variances of the heat and work, we can use the correlation functions to directly assess the temporal correlations of the current through the junction and the heat absorption and emission events. While the convergence of the density matrix to a constant steady state may be indicative of constant values of these quantities and their rates with no temporal structure, each discrete detection of a quantum of energy released or absorbed by the baths causes a measurement back action, i.e., a quench of the state of the engine. In the time immediately following the emission of a quantum into the cold bath, energy left in the hot qubit will start to oscillate between the qubit components  with frequency $E_J$. Only then will subsequent emission in the cold bath become possible. The periodic exchange of excitation between the qubits is eventually damped (with stronger damping as $\kappa$ increases), but correlations between emission events, as represented by $G^2(t,t')$,
may show a discernible temporal modulation, see Fig.~\ref{fig:corrbathsourkl2}(a). The back action from listening to the engine breaks the time symmetry and induces the characteristic chuffing enforced by the periodic motion of the piston in the classical steam engine.

To further address the function and regularity of the engine strokes we have additionally calculated the two-time correlations between the heat extraction from the hot bath and the current through the junction, $\langle  \hat{\sigma}_h^- (t) \hat{I} (t+\tau) \hat{\sigma}_h^+ (t) \rangle_{ss}$, between the current through the junction and the heat emission into the cold bath, $ \langle  \hat{I} (t) \hat{\sigma}_c^+ (t+\tau) \hat{\sigma}_c^- (t+\tau) \rangle_{ss}$, and between the heat extraction from the hot bath and the heat emission into the cold bath, $ \langle \hat{\sigma}_h^- (t) \hat{\sigma}_c^+ (t+\tau) \hat{\sigma}_c^- (t+\tau) \hat{\sigma}_h^+ (t) \rangle_{ss}$ (see Figs.~\ref{fig:corrbathsourkl2}(b),~\ref{fig:corrbathsourkl2}(c) and~\ref{fig:corrbathsourkl2}(d). \\
\indent It is interesting that while the density matrix is constant in steady state, the correlation functions clearly reveal how the incoherent excitation of the hot qubit by absorption from the hot bath, is followed by a transient transfer of energy towards the cold qubit and hence a positive amount of work. In the figure panels (b) and (c)  we can follow the oscillation of the quantum of energy, causing a later return of excitation with a negative production of work. The long time uncorrelated mean value of the work is positive on average.
%
\paragraph{Conclusion.} We have studied a simple autonomous two-qubit thermoelectric engine which allows us to calculate the work and heat mean values and fluctuations. We showed that even though those variables are very different in nature, their mean values and fluctuations are equivalent, i.e., for every heat quantum $\hbar \omega_c$ transferred to the cold heat bath a work of $2eV$ is done by the engine. These are quantum processes, and for the unobserved system it is not possible to assign strokes to the engine, but by calculating two-time correlation functions we predict a regularity in the  timing of quanta arriving into or leaving from the baths, and we predict bursts of the current around such detection events.
Our theoretical analysis addresses two crucial elements of quantum heat engines: the calculation of the efficiency and power of engines operating at the quantum level, and an assessment of fluctuations of work and heat. Using methods from quantum optics and open quantum systems we have shown how the fluctuations of work and heat can be defined and evaluated for autonomous systems with a constant steady state. These methods can be applied to a range of engine designs, and they may inspire efforts to quantify the interplay between, e.g., maximum power and maximum efficiency of heat engines and thermodynamic uncertainty relations~\cite{powertradeoff-seifert,powertradeoff-tasaki,powercriticalengine-campisi,efficiencybound-casati,uncertainty-fluctuations}.\\

The authors acknowledge financial support from the Villum Foundation.
%

\bibliography{Engine_manuscript_KV_210120_arXiv}
\end{document}